\documentclass[aps,pra,showpacs,twocolumn]{revtex4}
\usepackage{amsmath,graphicx,epsfig,amssymb,comment}

\newcommand{\tr}{\text{tr}}
\newcommand{\ten}{\otimes}
\newcommand{\ket}[1]{|#1\rangle}
\newcommand{\bra}[1]{\langle#1|}
\newcommand{\ketbra}[1]{\ket{#1}\bra{#1}}
\newcommand{\Ketbra}[2]{\ketbra{#1}\ten\ketbra{#2}}
\newcommand{\unit}{\mathbf{1}}
\newcommand{\half}{\frac{1}{2}}
\newcommand{\zero}{\mathbf{0}}
\newcommand{\SLHV}{\mathcal{S}_\text{LHV}}
\newcommand{\Sq}{\mathcal{S}_\text{QM}}
\newcommand{\Bell}{\mathcal{B}}

\newcommand{\eBellCH}{\langle\mathcal{B}_{\rm CH}\rangle}
\newcommand{\eBellCHSH}{\langle\mathcal{B}_{\rm CHSH}\rangle}

\newcommand{\bfx}{\mathbf{x}}
\newcommand{\bfy}{\mathbf{y}}
\newcommand{\bfw}{\mathbf{w}}
\newcommand{\bfr}{\mathbf{r}}

\newcommand{\vvec}[1]{\text{vec}(#1)}
\newcommand{\ibid}{{\em ibid.~}}

\begin{document}

\title{Bounds on Quantum Correlations in Bell Inequality Experiments}

\author{Yeong-Cherng~Liang}
\email{ycliang@physics.uq.edu.au}

\author{Andrew~C.~Doherty}
\email{doherty@physics.uq.edu.au} \affiliation{School of Physical
Sciences, The University of Queensland, Queensland 4072, Australia.}

\date{\today}

\begin{abstract}
Bell-inequality violation is one of the most widely known
manifestations of entanglement in quantum mechanics; indicating that
experiments on physically separated quantum mechanical systems
cannot be given a local realistic description. However, despite the
importance of Bell inequalities, it is not known in general how to
determine whether a given entangled state will violate a Bell
inequality. This is because one can choose to make many different
measurements on a quantum system to test any given Bell inequality
and the optimization over measurements is a high-dimensional
variational problem. In order to better understand this problem we
present algorithms that provide, for a given quantum state and Bell
inequality, both a lower bound and an upper bound on the maximal 
violation of the inequality. In many cases these bounds determine
measurements that would demonstrate violation of the Bell inequality
or provide a bound that rules out the possibility of a violation. Both
bounds apply techniques from convex optimization and the methodology
for creating upper bounds allows them to be systematically improved.
Examples are given to illustrate how these algorithms can be used to
conclude definitively if some quantum states violate a given Bell
inequality.
\end{abstract}

\maketitle

\section{Introduction}

The nature of entanglement has always been bewildering ever since
its first appearance in the
literature~\cite{EPR:1935,Schroedinger:Naturwissenschaften}. This is
more so after Bell's seminal work in 1964~\cite{J.S.Bell:1964}, in
which he showed, using what is now known as a Bell inequality, that
some experimental statistics of the spin-singlet state are
intrinsically incompatible with local realism.

For a long time after that, it seems to have been generally assumed
that entanglement and the violation of Bell inequalities are
synonymous. The first counterexample to that commonly held intuition
was provided by Werner~\cite{R.F.Werner:PRA:1989}, where he showed
that there are bipartite mixed two-qudit states, now known as the
Werner states, that are entangled and yet do not violate a large
class of Bell inequalities.

Soon after that, it was demonstrated by
Gisin~\cite{N.Gisin:PLA:1991} and later by Gisin and
Peres~\cite{N.Gisin:PLA:1992} that all bipartite pure entangled
states violate the Bell-Clauser-Horne-Shimony-Holt
inequality~\cite{Bell:CHSH}; the generalization to multipartite pure
entangled states was also carried out by Popescu and Rohrlich by
invoking appropriate postselection~\cite{S.Popescu:PLA:1992}. Three
years later, Horodecki {\em et~al.} provided the first analytic
criterion~\cite{RPM.Horodecki:PLA:1995} to determine if a two-qubit
state violates the Bell-Clauser-Horne-Shimony-Holt (henceforth
abbreviated as Bell-CHSH) inequality.

The Horodecki criterion is, unfortunately, also the only analytic
criterion that we have in determining if a broad class of quantum
states, namely two-qubit states, can be simulated classically. For
specific quantum states, there are examples where explicit local
hidden variables (LHV) models  have been constructed to reproduce
the quantum mechanical predictions, thereby ruling out the
possibility that these quantum states may violate a Bell
inequality~\cite{R.F.Werner:PRA:1989,LHV}.

In general, however, to determine if a quantum state violates a Bell
inequality is a high-dimensional variational problem, which requires
a nontrivial optimization of a Hermitian operator $\Bell$ (now known
as the {\em Bell} operator~\cite{S.L.Braunstein:PRL:1992}) over the
various possible measurement settings that each observer may
perform. This optimization does not appear to be convex and is
possibly $NP$-hard~\cite{M.M.Deza:Book:1997}.

Except for the simplest scenario  where one deals with Bell-CHSH
inequality~\cite{Bell:CHSH}, in conjunction with a two-qubit
state~\cite{RPM.Horodecki:PLA:1995}, or a maximally entangled pure
state~\cite{N.Gisin:PLA:1992,S.Popescu:PLA:1992b}, and its mixture
with the completely mixed state~\cite{T.Ito:PRA:2006}, very few
analytic results for the optimal measurements are known. As such,
for the purpose of characterizing quantum states that are
incompatible with local realistic description, efficient algorithms
to perform this state-dependent optimization are very desirable.

Bell-inequality violation is also relevant in various aspects of 
quantum information processing, in particular, quantum
teleportation~\cite{S.Popescu:PRL:1994}, quantum key
distribution~\cite{QKD1,QKD2}, and reduction of communication
complexity~\cite{Complexity}. Recently, it has even been
argued~\cite{QKD2} that Bell-inequality violation is necessary to
guarantee the security of some entanglement-based quantum key
distribution protocols.

On the other hand, state-independent bounds of quantum correlations
have also been investigated since the early 1980's. In particular,
Tsirelson~\cite{B.S.Tsirelson:LMP:1980} has demonstrated, using what
is now known as Tsirelson's vector construction, that in a Bell-CHSH
setup, bipartite quantum systems of arbitrary dimensions cannot
exhibit correlations stronger than $2\sqrt{2}$ - a value now known
as Tsirelson's bound. Recently, analogous bounds for more
complicated Bell inequalities have also been investigated by Filipp
and Svozil~\cite{S.Filipp:PRL:2004}, Buhrman and
Massar~\cite{H.Buhrman:PRA:2005}, Wehner~\cite{S.Wehner:PRA:2006},
Toner~\cite{B.F.Toner:quant-ph:0601172}, Avis {\em
et~al.}~\cite{D.Avis:JPA:2006} and Navascu\'es {\em
et~al.}~\cite{M.Navascues:quant-ph:0607119}. On a related note,
bounds on quantum correlations for given local measurements, rather
than given quantum state, have also been investigated by
Cabello~\cite{A.Cabello:PRL:2004} and Bovino {\em
et~al.}~\cite{F.A.Bovino:PRL:2004}.

In this paper, we will present, respectively, in
Sec.~\ref{sec:lower-bound} and Sec.~\ref{sec:upper-bound}, two
algorithms that were developed to provide a lower bound and an upper
bound on the maximal expectation value of a Bell operator for a
given quantum state. The second algorithm is another instance where
a nonlinear optimization problem is approximated by a hierarchy of
semidefinite programs, each giving a better bound of the original
optimization problem~\cite{M.Navascues:quant-ph:0607119,
P.A.Parrilo:2003,J.B.Lasserre:SJO:2001,hierarchy}. In its simplest
form, it provides a bound that is apparently state-independent, and
thus provides a (not necessarily tight) bound on the maximum
attainable quantum correlations in a Bell inequality experiment.

In Sec.~\ref{sec:examples}, we will derive, based on the second
algorithm, a necessary condition for a class of two-qudit states to
violate the Bell-CHSH inequality. We will also demonstrate how the
two algorithms can be used in tandem to determine if some quantum
states violate a Bell inequality. Some limitations of these
algorithms will then be discussed. We will conclude with a summary
of results and some possibilities for future research.

Throughout, boldfaced Latin letters, e.g., $\bfx$ will be used to
denote a column vector whereas $\zero$ and $\unit$, respectively,
represents a zero block matrix and an identity matrix. Moreover, the
$(i,j)$ entry of a matrix $M$ will be denoted by $[M]_{ij}$.

\section{Bounds on Quantum Correlations}\label{sec:measurements}

\subsection{Preliminaries}

Bell inequalities are inequalities derived from the assumption of a
general LHV model. A particular Bell inequality deals with a
specific experimental setup, say a source that distributes pairs of
particles to two experimenters (hereafter called Alice and Bob), and
where each of them can perform, respectively, $m_A$ and $m_B$
alternative measurements that would each generate $n_A$ and $n_B$
distinct outcomes. For each of these setups, a Bell inequality
places a bound on the experimental statistics obtained from the
corresponding Bell experiments. If there exists a LHV model that
saturates the bound then the inequality is said to be
tight~\cite{fn:tight inequality}. In what follows, we will adopt the
notation introduced in Ref.~\cite{D.Collins:JPA:2004} and refer to a
tight Bell inequality for such an experimental setup as a
Bell-$m_Am_Bn_An_B$ inequality.

A Bell inequality for correlations, such as the Bell-CHSH
inequality~\cite{Bell:CHSH} typically involves statistical
constraints on some linear combination of {\em correlation
functions}. Similarly, a Bell inequality for probabilities, such as
the Bell-Clauser-Horne (henceforth abbreviated as Bell-CH)
inequality~\cite{CH:PRD:1974}, places bounds on some linear
combinations of joint and marginal {\em probabilities} of
experimental outcomes. In either case, a general Bell inequality
takes the form:
\begin{equation}
\mathcal{S}_{\rm LHV}\le \beta_{\rm LHV},\label{Eq:Scl:dfn}
\end{equation}
where $\beta_{\rm LHV}$ is a real number and $\mathcal{S}_{\rm LHV}$
involves a specific linear combination of correlation functions or
joint and marginal probabilities of experimental outcomes.

To compare with predictions given by quantum mechanics, these
correlation functions, or probabilities, are calculated using the
quantum mechanical rules. The bounds on $\SLHV$ then translate into
corresponding bounds $\beta_{\rm LHV}$ on the expectation value of
some Hermitian observable that describes the Bell inequality
experiment, this observable is known as the {\it Bell operator}
$\mathcal{B}$~\cite{S.L.Braunstein:PRL:1992}. The restriction that
the given Bell inequality is satisfied in the experiment is then
\begin{equation}
\Sq(\rho,\Bell) = \tr\left(\rho~\mathcal{B}\right)\le \beta_{\rm
LHV}.\label{Eq:S:dfn}
\end{equation}
The Bell operator depends on the choice of measurements at each of
the sites (polarizer angles for example). These measurements will be
described by a set of Hermitian operators $\{O_m\}$. For correlation
inequalities these are simply the measured observables at each stage
of the Bell measurement, while for general probability inequalities
the $O_m$ are elements of the positive-operator-valued measures
(POVMs) that describe the measurements at each site. We will denote
this expectation value by $\Sq(\rho, \{O_m\})$ when we want to
emphasize its dependence on the choice of local Hermitian
observables $O_m$. Ideally the choice of measurement should give the
maximal expectation value of the Bell operator, for which we will
give the notation
\begin{equation}
\Sq(\rho)\equiv\max_{\{O_m\}}\Sq(\rho,\{O_m\}).
\end{equation}
It is this implicitly-defined function that will give us information
about which states violate a given Bell inequality.

As an example, let us recall the Bell-CHSH
inequality~\cite{Bell:CHSH}, which is a dichotomic (i.e.,
two-outcome) Bell correlation  inequality that involves two parties,
and where the two possible measurement outcomes are assigned the
values $\pm1$:
\begin{align}
\SLHV&=E(A_1,B_1)+E(A_1,B_2)+E(A_2,B_1)-E(A_2,B_2)\nonumber\\
&\le 2.\label{Eq:Bell-CHSH-inequality}
\end{align}
In the above expression, the correlation function $E(A_k,B_l)$
represents the expectation value  of Alice's measurement outcome
times Bob's measurement outcome, given that she has chosen to
measure the observable $A_k$ and he has chosen to measure the
observable $B_l$. In quantum mechanics, these correlation functions
are computed using
\begin{equation}
E(A_k,B_l)=\tr\left(\rho~A_k\ten B_l\right).
\end{equation}
Substituting this into Eq.~\eqref{Eq:Bell-CHSH-inequality} and
comparing with Eq.~\eqref{Eq:S:dfn}, one finds that the
corresponding Bell operator reads
\begin{equation}\label{Eq:Bell-CHSH:operator}
\Bell_\text{CHSH}=A_1\ten(B_1+B_2)+A_2\ten(B_1-B_2).
\end{equation}

To determine the maximal Bell-inequality violation for a given
$\rho$, $\Sq(\rho)$,  requires a maximization by varying over all
possible choices of $\{O_m\}$, i.e., $A_k$ and $B_l$ in the case of
Eq.~\eqref{Eq:Bell-CHSH:operator}.  Whether we are interested in
correlation inequalities or in Bell inequalities for probabilities
the (bipartite) Bell operator has the general structure
\begin{equation}\label{Eq:BellOperator:General}
\mathcal{B}=\sum_{K,L}b_{KL} A_K\ten
B_L.
\end{equation}
In the case of a Bell inequality for probabilities the indices $K,L$
are collective indices describing both a particular measurement
setting and a particular outcome for each observer. For correlation
inequalities they refer simply to the measurement settings as in the
Bell-CHSH case described in detail above.

In what follows, we will present two algorithms which we have
developed specifically to perform the maximization over choice of
measurements. The first, which we will abbreviate as LB, provides a
{\em lower bound} on the maximal expectation value and can be
implemented for any Bell inequality. This bound makes use of the
fact that the objective function $\Sq$ is bilinear in the
observables $O_m$, that is it is linear in the $A_K$ for fixed $B_L$
and likewise linear in the $B_L$ for fixed $A_K$. The second bound,
which we will abbreviate as UB, provides an {\em upper bound} on
$\Sq(\rho)$ by regarding $\Sq(\rho,\{O_m\})$ as a polynomial
function of the variables that define the various $O_m$ and applying
general techniques for finding such bounds on
polynomials~\cite{P.A.Parrilo:2003,J.B.Lasserre:SJO:2001}.

Both of these make use of convex optimization techniques in  the
form of a semidefinite program
(SDP)~\cite{L.Vandenberghe:SR:1996,S.Boyd:Book:2004}. A semidefinite
program is an optimization over Hermitian matrices. The objective
function depends linearly on the matrix variable (as expectation
values do in quantum mechanics for example) and the optimization is
carried out subjected to the constraint that the matrix variable is
positive semidefinite and satisfies various affine constraints. Any
semidefinite program may be written in the following {\it standard
form}:
\begin{subequations}\label{Eq:SDP:Matrix}
\begin{align}
&\text{maximize \ }-\text{tr}\left[ F_{0}Z\right], \label{Eq:SDP:Matrix:Obj}\\
&\text{subject to\ }\quad\tr\left[ F_{m}Z\right] =c_{m} \quad \forall~m, \label{Eq:SDP:Matrix:Eq}\\
&\text{ \ \ \ \ \ \ \ \ \ \ \ \ \ \ \ \ \ \ \ \ }Z\geq
0,\label{Eq:SDP:Matrix:Ineq}
\end{align}\end{subequations}
where $F_0$ and all the $F_m$'s are Hermitian matrices and the $c_m$
are real numbers that together specify the optimization; $Z$ is
the Hermitian matrix variable to be optimized.

An SDP also arises naturally in the {\em inequality form}, which
seeks  to minimize a linear function of the optimization variables
$\bfx\in\mathbb{R}^n$, subjected to a linear matrix inequality:
\begin{subequations}\label{Eq:SDP:Vector}
\begin{gather}
\text{minimize \ \ \ \ }\quad \mathbf{x}^T\mathbf{c'}\qquad\label{Eq:SDP:Vector:Obj}\\
\text{subject to \ \ }G_0+\sum_{m}\bfx_mG_m
\geq 0.\label{Eq:SDP:Vector:Ineq}
\end{gather}\end{subequations}
As in the standard form, $G_0$ and all the $G_m$'s are Hermitian
matrices, while $\mathbf{c'}$ is a real vector of length $n$.

\subsection{Lower Bound on $\Sq(\rho)$}\label{sec:lower-bound}
The key idea behind the LB algorithm is to realize that when
measurements for all but one party are {\em fixed}, the optimal
measurements for the remaining party can be obtained efficiently
using convex optimization techniques, in particular an SDP. Thus we
can fix Bob's measurements and find Alice's optimal choice, at least
numerically; with this optimized measurements for Alice, we can
further find the optimal measurements for Bob (for this choice of
Alice's settings), and then Alice again and so on and so forth until
$\Sq(\rho,\{O_m\})$ converges within the desired numerical
precision~\cite{fn:T.Ito}.

Back in 2001, Werner and Wolf~\cite{R.F.Werner:QIC:2001} presented a
similar iterative algorithm, by the name of {\em See-Saw iteration},
to maximize the expectation value of the Bell operator for a
correlation inequality involving only dichotomic
observables~\cite{fn:Dichotomic:Defn}. As a result we will focus
here on the (straightforward) generalization to the widest possible
class of Bell inequalities. In the work of Werner and
Wolf~\cite{R.F.Werner:QIC:2001} it turned out that once the
dichotomic observables for one party are fixed, optimization of the
other party's observables can be carried out explicitly. This turns
out to be true for any dichotomic Bell inequality and we will
return to this question in Sec.~\ref{sec:two-outcome}.

\subsubsection{General Settings}
Firstly we must develop a more explicit notation for a general Bell
inequality for probabilities~\cite{fn:probability-correlation}. Let
us consider a Bell-$m_Am_Bn_An_B$ inequality for probabilities. We
will denote the POVM element associated with the $\kappa^\text{th}$
outcome of Alice's $k^\text{th}$ measurement by $A^\kappa_k$ while
$B^\lambda_l$ is the POVM element associated with the
$\lambda^\text{th}$ outcome of Bob's $l^\text{th}$ measurement.
Moreover, let $d_A$ and $d_B$, respectively, be the dimension of the
state space that each of the $A^\kappa_k$ and $B^\lambda_l$ acts on.
Then it follows from Born's rule
that\begin{subequations}\label{Eq:QM:probabilities}
\begin{gather}
p^{\kappa\lambda}_{AB}(k,l)= \tr\left(\rho\,A^\kappa_k\ten
B^\lambda_l\right)\label{Eq:QM:probabilities:joint}\\
p^\kappa_A(k)=\tr\left(\rho\,A^\kappa_k\ten \unit_{d_B}\right),\quad
p^\lambda_B(l)= \tr\left(\rho\,\unit_{d_A}\ten
B^\lambda_l\right),\label{Eq:QM:probabilities:marginal}
\end{gather}\end{subequations}
where $p^{\kappa\lambda}_{AB}(k,l)$ refers to the joint probability
that the $\kappa^{\rm th}$ experimental outcome is observed at
Alice's site and the $\lambda^{\rm th}$ outcome at Bob's, given that
Alice performs the $k^{\rm th}$ and Bob performs the $l^{\rm th}$
measurement. The marginal probabilities $p^\kappa_A(k)$ and
$p^\lambda_B(l)$ are similarly defined. A general Bell operator for
probabilities can then be expressed as
\begin{equation}
\mathcal{B}=\sum_{k=1}^{m_A}\sum_{\kappa=1}^{n_A}\sum_{l=1}^{m_B}
\sum_{\lambda=1}^{n_B}b^{\kappa\lambda}_{kl} A^\kappa_k\ten
B^\lambda_l,\label{Eq:BellOperator}
\end{equation}
where $b^{\kappa\lambda}_{kl}$ are determined from the given Bell
inequality. Note that the sets of POVM elements
$\left\{A^\kappa_k\right\}_{\kappa=1}^{n_A}$ and
$\left\{B^\lambda_l\right\}_{\lambda=1}^{n_B}$ satisfy
\begin{subequations}\label{Eq:POVM}
\begin{gather} \sum_{\kappa=1}^{n_A}A^\kappa_k=\unit_{d_A}\quad
\text{and}\quad\sum_{\lambda=1}^{n_B}B^\lambda_l=\unit_{d_B}
\quad\forall\quad k, l,\label{Eq:POVM:Normalization}\\
 A^\kappa_k\ge 0,\quad B^\lambda_l\ge
0\quad\forall\quad k, l,\kappa,\lambda.\label{Eq:POVM:PSD}
\end{gather}
\end{subequations}

\subsubsection{Iterative Semidefinite Programming Algorithm}

To see how to develop a lower bound on $\Sq(\rho)$ by fixing the
observables at one site and optimizing the other, we observe that
upon substituting Eq.~\eqref{Eq:BellOperator} into
Eq.~\eqref{Eq:S:dfn}, the  left-hand-side of the inequality can be
rewritten as
\begin{align}
\Sq(\rho,A^\kappa_k,B^\lambda_l)&=\sum_{l,\lambda}
\tr \left(\rho_{B^\lambda_l}\,B^\lambda_l\right),\label{Eq:S2}
\end{align}
where $\rho_{B^\lambda_l}\equiv\sum_{k,\kappa}b^{\kappa\lambda}_{kl}~
\tr_A\left[\rho\left(A^\kappa_k\ten\unit_{d_B}\right)\right]$.
$\tr_A~\cdot$ is the partial trace over subsystem $A$.

Notice that if $\rho_B\,^\lambda_l$ are held constant by fixing all
of Alice's measurement settings (given by the set of $A^\kappa_k$)
then  $\rho_B\,^\lambda_l$ is a constant matrix independent of the
$B^\lambda_l$. Thus the objective function is linear in these
variables. The constraints that $\{B^\lambda_l\}_{\lambda=1}^{n_B}$
form a POVM for each value of $l$ is a combination of affine and
matrix nonnegativity constraints. As a result it is fairly clear
that the following problem is an SDP in standard form
\begin{subequations}\label{Eq:sdpiter}
\begin{eqnarray}
&{\rm maximize}_{\{B^\lambda_l\}}&\Sq(\rho,A^\kappa_k,B^\lambda_l)\label{Eq:SDPIter:obj}\\
&{\rm subject \ to} \quad &
\sum_{\lambda=1}^{n_B}B^\lambda_l=\unit_{d_B}
\quad\forall\quad l, \label{Eq:SDPIter:eq}\\
 && B^\lambda_l\ge
0\quad\forall\quad l, \lambda \label{Eq:SDPIter:ineq}
\end{eqnarray}
\end{subequations}
Explicit forms for the matrices $F_m$ and values $c_m$ of
Eq.~\eqref{Eq:SDP:Matrix} can be found in Appendix~\ref{sec:SDP}.

Exactly the same analysis follows if we fix Bob's measurement
settings, and optimize over Alice's POVM elements instead. To arrive
at a local maximum of $\Sq(\rho,A^\kappa_k,B^\lambda_l)$, it
therefore suffices to start with some random measurement settings
for Alice (or Bob), and optimize over the two parties' settings
iteratively. A (nontrivial) lower bound on $\Sq(\rho)$ can then be
obtained by optimizing the measurement settings starting from a set
of randomly generated initial guesses.

It is worth noting that in any implementation of this algorithm,
physical observables $\{A^\kappa_k, B^\lambda_l\}$ achieving the lower
bound are constructed when the corresponding SDP is solved. In the
event that the lower bound is greater than the  
classical threshold $\beta_\text{LHV}$, then these observables can, in
principle, be measured in the laboratory to demonstrate a
Bell-inequality violation of the given quantum state.

We have implemented this algorithm in MATLAB~\cite{fn:MATLAB} to
search for a lower bound on $\Sq(\rho)$ in the case of
Bell-CH~\cite{CH:PRD:1974}, $I_{3322}$, $I_{4422}$, $I_{2233}$ and
$I_{2244}$ inequality~\cite{D.Collins:JPA:2004}, and with the local
dimension $d$ up till 32. Typically, with no more than 50
iterations, the algorithm already converges to a point that is
different from a local maximum by no more than $10^{-9}$. To test
against the effectiveness of finding $\Sq(\rho)$ using LB, we have
randomly generated 200 Bell-CH violating two-qubit states and found
that on average, it takes about 6 random initial guesses before the
algorithm gives $\Sq(\rho,\{O_m\})$ that is close to the actual
maximum (computed using Horodecki's
criterion~\cite{RPM.Horodecki:PLA:1995} within $10^{-5}$. Specific
examples regarding the implementation of this algorithm will be
discussed in Sec.~\ref{sec:examples}.

Two other remarks concerning this algorithm should now be made.
Firstly, the algorithm is readily generalized to multipartite Bell
inequalities for probabilities: one again starts with some random
measurement settings for all but one party, and optimizes over each
party iteratively. Also, it is worth noting that this algorithm is
not only useful as a numerical tool, but for specific cases, it can
also provide useful analytic criterion. In particular, when applied
to the Bell-CH inequality~\cite{CH:PRD:1974} for two-qubit states,
the analysis for dichotomic observables discussed in the next
subsection allows one to recover the Horodecki's
criterion~\cite{RPM.Horodecki:PLA:1995}, i.e.,~the necessary and
sufficient condition for two-qubit states to violate the Bell-CHSH
inequality~\cite{Bell:CHSH,fn:CH-CHSH}.

\subsubsection{Two-outcome Bell Experiment}\label{sec:two-outcome}

We will show that, just as in the case of correlation
inequalities~\cite{R.F.Werner:QIC:2001}, the local optimization can
be solved analytically for two-outcome measurements. If we denote by
``$\pm$" the two outcomes of the experiments, it follows from
Eq.~\eqref{Eq:POVM} that the POVM element $B^-_l$ can be expressed
as a function of the complementary POVM element $B^+_l$,
i.e.,~$B^-_l=\unit_{d_B}-B^+_l$, subjected to $0\le B^+_l \le
\unit_{d_B}$. We then have
\begin{align*}
\sum_{\lambda=\pm}\tr \left(\rho_{B^\lambda_l}\,B^\lambda_l\right)&=
\tr \left[\left(\rho_{B^+_l}-\rho_{B^-_l}\right)B^+_l\right] +\tr
\left(\rho_{B^-_l}\right).
\end{align*}
The above expression can be maximized by setting the positive
semidefinite operator $B^+_l$ to be the projector onto the positive
eigenspace of $\rho_{B^+_l}-\rho_{B^-_l}$. In a similar manner, we
can also write
\begin{align*}
\sum_{\lambda=\pm}\tr
\left(\rho_{B^\lambda_l}\,B^\lambda_l\right)=\tr
\left[\left(\rho_{B^-_l}-\rho_{B^+_l}\right)B^-_l\right]+\tr
\left(\rho_{B^+_l}\right),
\end{align*}
which can be maximized by setting  $B^-_l$ to be the projector onto
the nonpositive eigenspace of $\rho_{B^+_l}-\rho_{B^-_l}$. Notice
that this choice is consistent with our earlier choice of $B\,^+_l$
for the $+$ outcome POVM element in that they forms a valid POVM.
Since there can be no difference in these maxima, we may write the
maximum as their average, i.e.,~
\begin{align*}
\sum_{\lambda=\pm}\tr
\left(\rho_{B^\lambda_l}\,B^\lambda_l\right)&=\half\left|\left|
\rho_{B^+_l}-\rho_{B^-_l}\right|\right|+\half\sum_{\lambda=\pm}\tr
\left(\rho_{B^\lambda_l}\right),
\end{align*}
where $||O||$ is the trace norm of the Hermitian operator
$O$~\cite{LA}. Carrying out the optimization for each of the $l$
settings, the optimized $\Sq(\rho,A^\kappa_k,B^\lambda_l)$, as an
implicit function  of  Alice's POVM $\left\{A^\kappa_k\right\}$, is
given by
\begin{align}
\Sq(\rho,A^\kappa_k)&=\half\sum_l\left|\left|\rho_{B^+_l}-\rho_{B^-_l}\right|\right| 
+\half\sum_l\sum_{\lambda=\pm}\tr
\left(\rho_{B^\lambda_l}\right).\label{Eq:S:implicitA}
\end{align}

An immediate corollary of the above result is that for the
optimization of a two-outcome Bell operator for probabilities, it is
unnecessary for any of the two observers to perform generalized
measurements described by a POVM; von Neumann projective
measurements are sufficient~\cite{fn:sufficiency.projector}. In
practice, this simplifies any analytic treatment of the optimization
problem as a generic parametrization of a POVM is a lot more
difficult to deal with, thereby supporting the simplification
adopted in Ref.~\cite{S.Filipp:PRL:2004}.

Nevertheless, it may still be advantageous to consider generic POVMs
as our initial measurement settings when implementing the algorithm
numerically. This is because the local maximum of $\Sq(\rho,\{O_m\})$
obtained using the iterative procedure is a function of
the initial guess. In particular, it was found that the set of local
maxima attainable could change significantly if the ranks of the
initial measurement projectors are altered. As such, it seems
necessary to step through various ranks of the starting projectors
to obtain a good lower bound on $\Sq(\rho)$. Even then, we have also
found examples where this does not give a lower bound on $\Sq(\rho)$
that is as good as when generic POVMs are used as the initial
measurement operators.

\subsection{Upper Bound on $\Sq(\rho)$}\label{sec:upper-bound}
A major drawback of the above algorithm, or the analogous algorithm
developed by Werner and Wolf~\cite{R.F.Werner:QIC:2001} for Bell
correlation inequalities is that, except in some special cases, it
is generally impossible to tell if the maximal $\Sq(\rho,\{O_m\})$
obtained through this optimization procedure corresponds to the
global maximum $\Sq(\rho)$.

Nontrivial upper bounds on $\Sq(\rho)$, nevertheless, can be
obtained by considering {\em relaxations} of the global optimization
problem. In a relaxation, a (possibly nonconvex) maximization
problem is modified in some way so as to yield a more tractable
optimization that bounds the optimization of interest. One example
of a variational upper bound that exists for any optimization
problem is the {\em Lagrange dual} optimization that arises in the
method of Lagrange multipliers~\cite{S.Boyd:Book:2004}.

To see how to apply existing studies in the optimization literature
to find upper bounds on $\Sq(\rho)$, let us first remark that the
global objective function $\Sq(\rho,\{O_m\})$ can be mapped to a
polynomial function in real variables, for instance, by expanding
all the local observables $\{O_m\}$ and the density matrix $\rho$ in
terms of Hermitian basis operators.  In the same manner, matrix
equality constraints, such as that given in
Eq.~\eqref{Eq:POVM:Normalization} can also mapped to a set of
polynomial equalities by requiring that the matrix equality holds
component wise. Now, it is known that a hierarchy of global bounds
of a polynomial function, subjected to polynomial equalities and
inequalities, can be achieved by solving suitable
SDPs~\cite{P.A.Parrilo:2003,J.B.Lasserre:SJO:2001}. Essentially,
this is achieved by approximating the original nonconvex
optimization problem by a series of convex ones in the form of a
SDP, each giving a better bound of the original polynomial objective
function.

At the bottom of this hierarchy is the lowest order relaxation
provided by the {\em Lagrange dual} of the original nonconvex
problem. By considering {\em Lagrange multipliers} that depend on
the original optimization variables, higher order relaxations to the
original problem can be constructed to give tighter upper bounds on
$\Sq(\rho)$ (see Appendix~\ref{sec:semidefinite relaxation} for more
details).

In the following, we will focus our discussion on a general
two-outcome Bell (correlation) inequality, where the observables
$\{O_m\}$ are only subjected to matrix equalities. In particular, we
will show that the global optimization problem for these Bell
inequalities is a {\em quadratically-constrained quadratic-program}
(QCQP), i.e., one whereby the objective function and the constraints
are both {\em quadratic} in the optimization variables. Then, we
will demonstrate explicitly how the Lagrange dual of this QCQP,
which is known to be an SDP, can be constructed. The analogous
analytic treatment is apparently formidable for higher order
relaxations. Nonetheless, there exists third-party
MATLAB~\cite{fn:MATLAB} toolbox known as the SOSTOOLS which is
tailored specifically for this kind of optimization
problem~\cite{SOSTOOLS,S.Prajna:LNCIS:2005}.

Numerically, we have implemented the algorithm for several
two-outcome correlation inequalities and will discuss the results in
greater detail in Sec.~\ref{sec:examples}. For a general Bell
inequality where each $O_m$ is also subjected to a linear matrix
inequality (LMI) like Eq.~\eqref{Eq:POVM:PSD}, the algorithm can
still be implemented, for instance, by requiring that all the
principle submatrices of $O_m$ have nonnegative
determinants~\cite{LA}. This then translates into a set of
polynomial inequalities which fit into the framework of a general
polynomial optimization problem (see Appendix~\ref{sec:semidefinite
relaxation}). However a more effective approach would retain the
structure of linear matrix inequalities constraining a polynomial
optimization problem; we leave the investigation of these bounds to
further work.

\subsubsection{Global Optimization Problem}\label{sec:GlobalOptimizationProblem}

Now, let us consider a dichotomic Bell correlation inequality where
Alice an Bob can respectively perform $m_A$ and $m_B$ alternative
measurements. A general Bell correlation operator for such an
experimental setup can be written as~\cite{fn:Bell-correlation}
\begin{equation}
\mathcal{B}=\sum_{k=1}^{m_A}\sum_{l=1}^{m_B} b_{kl} O_k\ten
O_{l+m_A},\label{Eq:BellOperator:Correlation1}
\end{equation}
where $b_{kl}$ are determined from the given Bell correlation
inequality, $O_k$ for $k=1,\ldots,m_A$ refers to the $k^{\rm th}$
Hermitian observable measured by Alice, and  $O_{l+m_A}$ for
$l=1,\ldots,m_B$ refers to the $l^{\rm th}$ Hermitian observable
measured by Bob. Furthermore, by convention, these dichotomic
observables can be chosen to have eigenvalues $\pm1$~\cite{fn:pm1}
and thus
\begin{gather}
O_m^\dag O_m=(O_m)^2=\unit_{d}\quad \forall\quad
m,\label{Eq:Constraint:Equality}
\end{gather}
where we have assumed for simplicity that all the local observables
$O_m$ act on a state space of dimension
$d$~\cite{fn:subsystems.diff.dim}.

The global optimization problem derived from a dichotomic Bell
correlation inequality thus takes the form of
\begin{subequations}\label{Eq:GlobalOptimization:Correlation}
\begin{align}
&\text{maximize \ \ \ } \tr\left(\rho~\Bell\right),\label{Eq:GlobalOptimization:Correlation:Obj}\\
&\text{subject to \ \ } O_m^2=\unit_d \quad
\forall~m=1,2,....\label{Eq:GlobalOptimization:Correlation:Constraint}
\end{align}\end{subequations}
For any $m\times n$ complex matrices, we will now define \vvec{$A$}
to be the $m\cdot n$ dimensional vector obtained by stacking all
columns of $A$ on top of one another. By collecting all the
vectorized observables together
\begin{equation*}
\bfw^\dag\equiv[\vvec{O_1}^\dag\,\,\vvec{O_2}^\dag\,\ldots\,\,\vvec{O_{m_A+m_B}}^\dag],
\end{equation*}
and using the identity
\begin{equation}
\tr(\rho~O_k\ten O_{l+m_A})=\vvec{O_k}^\dag(V\rho)^{T_A}\vvec{O_{l+m_A}},
\end{equation}
with $V$ being the flip operator such that
$V\ket{i}\ket{j}=\ket{j}\ket{i}$  and $(.)^{T_A}$ being
the partial transposition with respect to subsystem $A$,
we can write the objective function more explicitly as
\begin{equation}\label{Eq:eBell:Correlation:Poly}
\Sq(\rho,\{O_m\})=\tr(\rho~\Bell)=-\bfw^\dag\Omega_0\bfw
\end{equation}
where
\begin{equation*}
\Omega_0\equiv\half\left(
\begin{array}{cc}
\zero & -b\ten R \\
-b^T\ten R^\dag & \zero\\
\end{array}\right),
\end{equation*}
$b$ is a $m_A\times m_B$ matrix with $[b]_{kl}=b_{kl}$
[c.f.~Eq.\eqref{Eq:BellOperator:Correlation1}] and $R\equiv
(V\rho)^{T_A}$. In this form, it is explicit that the objective
function is {\em quadratic} in $\vvec{O_m}$. Similarly, by requiring
that the matrix equality holds component-wise, we can get a set of
equality constraints, which are each {\em quadratic} in
$\vvec{O_m}$. The global optimization problem
\eqref{Eq:GlobalOptimization:Correlation} is thus an instance of a
QCQP.

On a related note, for any Bell inequality experiments where
measurements are restricted to the projective type, the global
optimization problem is also a QCQP. To see this, we first note that
the global objective function for the general case, as follows from
Eq.~\eqref{Eq:S:dfn} and Eq.~\eqref{Eq:BellOperator:General}, is
always quadratic in the local Hermitian observables
$\{A_K,B_L\}$. The requirement that these measurement operators are
projectors amounts to requiring
\begin{align}
A_K^2=A_K,\quad B_L^2=B_L,\quad \forall~K,L,
\end{align}
which are quadratic constraints on the local Hermitian observables.
Since we have shown in Sec.~\ref{sec:two-outcome} that for any
two-outcome Bell inequality for probabilities, it suffices to
consider projective measurements in optimizing $\Sq(\rho,\{O_m\})$,
it follows that the global optimization problem for these Bell
inequalities is always a QCQP.

\subsubsection{State-independent Bound}\label{sec:State-independent Bound}

As mentioned above, the lowest order relaxation to the global
optimization problem \eqref{Eq:GlobalOptimization:Correlation} is
simply the Lagrange dual of the original QCQP. This can be obtained
via the {\em Lagrangian} of the global optimization problem
\eqref{Eq:GlobalOptimization:Correlation}
\begin{equation}
\mathcal{L}(\{O_m\},\Lambda_m)=\Sq(\rho,\{O_m\})-\sum_{m=1}^{m_A+m_B}
\tr\left[\Lambda_m\left(O_m^2-\unit_d\right)\right],
\label{Eq:Lagrangian}
\end{equation}
where $\Lambda_m$ is a matrix of Lagrange multipliers associated
with the $m^{\rm th}$ matrix equality constraint. With no loss of
generality, we can assume that $\Lambda_m$'s are Hermitian.

Notice that for all values of $\{O_m\}$ that satisfy the
constraints, the Lagrangian $\mathcal{L} (\rho, \{O_m\},\Lambda_m)=
\Sq(\rho,\{O_m\})$ . As a result, if we maximize the Lagrangian
without regard to the constraints we obtain an upper bound on the
maximal expectation value of the Bell operator
\begin{equation}
  \max_{\{O_m\}} \mathcal{L} (\rho, \{O_m\},\Lambda_m) \geq \Sq(\rho).
\end{equation}
The Lagrange dual optimization simply looks for the best such upper
bound.

In order to maximize the Lagrangian we rewrite the Lagrangian
with Eq.~\eqref{Eq:eBell:Correlation:Poly} and the identity
\begin{equation}
\tr\left(\Lambda_mO_mO_m^\dag\right)=\vvec{O_m}^\dag\unit_{d}\ten
\Lambda_m\vvec{O_m},
\end{equation}
to obtain
\begin{equation}
\mathcal{L}(\bfw,
\Lambda_m)=-\bfw^\dag\Omega\bfw+\sum_{m=1}^{m_A+m_B}\tr~\Lambda_m,
\label{Eq:Lagrangian:Explicit}
\end{equation}
where $\Omega\equiv\Omega_0+\bigoplus_{m=1}^{m_A+m_B}\unit_{d}\ten
\Lambda_m$. Note that each of the diagonal blocks $\unit_{d}\ten
\Lambda_m$ is of the same size as the matrix $R$.

To obtain the dual optimization problem, we maximize the Lagrangian
over  $\bfw$ to obtain the {\em Lagrange dual function}
\begin{equation}\label{Eq:Lagrange Dual}
g(\Lambda_m)\equiv \sup_{\bfw}\mathcal{L}(\bfw,\Lambda_m).
\end{equation}
As noted above $g(\Lambda_m)\ge \Sq(\rho)$ for all choices of
$\Lambda_m$. Moreover, this supremum  over $\bfw$ is unbounded above
unless $\Omega \ge 0$, in which case the supremum is attained by
setting $\bfw=0$ in Eq.~\eqref{Eq:Lagrangian:Explicit}. Hence, the
{\em Lagrange dual} optimization, which seeks for the best upper
bound of Eq.~\eqref{Eq:GlobalOptimization:Correlation} by minimizing
Eq~\eqref{Eq:Lagrange Dual} over the Lagrange multipliers, reads
\begin{gather}
\text{minimize \ \ } \sum_{i=1}^{m_A+m_B}\tr~\Lambda_m,\nonumber\\
\text{subject to \ \ \ }\quad\Omega \geq
0.\qquad\label{Eq:SDP-relaxation:deg2}
\end{gather}
By expanding $\Lambda_m$ in terms of Hermitian basis operators
satisfying Eq.~\eqref{Eq:basis:Gell-Mann},
\begin{equation}\label{Eq:Lambda:Expansion}
\Lambda_m=\sum_{n=0}^{d^2-1}\lambda_{mn}\sigma_n,
\end{equation}
the optimization problem~\eqref{Eq:SDP-relaxation:deg2} is readily
seen to be an SDP in the inequality form~\eqref{Eq:SDP:Vector}.

For Bell-CHSH inequality and the correlation equivalent of
$I_{3322}$ inequality, it was observed numerically that the upper
bound obtained via the SDP \eqref{Eq:SDP-relaxation:deg2} is always
{\em state-independent}. In fact, for 1000 randomly generated
two-qubit states, and 1000 randomly generated two-qutrit states, the
Bell-CHSH upper bound of $\Sq(\rho)$ obtained through
\eqref{Eq:SDP-relaxation:deg2} was never found to differ from the
Tsirelson bound~\cite{B.S.Tsirelson:LMP:1980} by more than
$10^{-7}$. In fact by finding an explicit feasible solution to the
optimization problem dual to Eq.~\eqref{Eq:SDP-relaxation:deg2},
Wehner has shown that the upper bound obtained here can be no better
than that obtained by Tsirelson's vector construction for
correlation inequalities~\cite{S.Wehner}.

In a similar manner, we have also investigated the upper bound of
$\Sq(\rho)$ for some dichotomic Bell probability inequalities using
the lowest order relaxation to the corresponding global optimization
problem. Interestingly, the numerical upper bounds obtained from the
analog of Eq.~\eqref{Eq:SDP-relaxation:deg2} for these inequalities,
namely the Bell-CH inequality, the $I_{3322}$ inequality and the
$I_{4422}$ inequality, are also found to be state-independent and
are given by 0.2071067, 0.375 and 0.6693461 respectively.

\subsubsection{State-dependent Bound}\label{sec:State-dependent Bound}

Although the state-independent upper bounds obtained above are
interesting in their own right, our main interest here is to find an
upper bound  on $\Sq(\rho)$ that does depend on the given quantum
state $\rho$. This can be obtained, with not much extra cost, from
the Lagrange dual to a more-refined version of the
original optimization problem.

To appreciate that, let us first recall that each dichotomic
Hermitian observable $O_m$ can only have eigenvalues $\pm1$. It
follows that their trace\begin{subequations}\label{Eq:z}
\begin{equation}
z_m\equiv\tr(O_m),\label{Eq:z_m:Defn}
\end{equation}
can only take on the following values
\begin{equation}
z_m=-d, -d+2,\ldots,d-2, d.\label{Eq:Range:z}
\end{equation}\end{subequations}
In particular, if $z_{m}=\pm d$ for any $m$, then $O_m=\pm\unit_d$
and it is not difficult to show that the Bell-CHSH inequality cannot
be violated for this choice of observable.

Better Lagrange dual bounds arise from taking these additional
constraints \eqref{Eq:z} explicitly into account. We found it most
convenient to express the original optimization problem in terms of
real variables given by the expansion coefficients of $O_m$ in terms
of a basis for Hermitian matrices that includes the (traceless)
Gell-Mann matrices and the identity matrix [c.f.
Eq.~\eqref{Eq:basis:Gell-Mann}]. For details see
Appendix~\ref{sec:fixed-rank}. The result is a set of SDPs, one for
each of the various choices of $z_m$. The lowest order upper bound
on $\Sq(\rho)$ can then be obtained by stepping through the various
choices of $z_m$ given in Eq.~\eqref{Eq:Range:z}, solving each of
the corresponding SDP, and taking their maximum. The results of this
approach will be discussed later, for now it suffices to note that
tighter bounds can be obtained which are explicitly state dependent.

\subsubsection{Higher Order Relaxations}
The higher order relaxations simply arise from allowing the Lagrange
multipliers $\lambda$ to be polynomial functions of $\{O_m\}$ rather
than constants. In this case, it is no longer possible to optimize
over the primal variables in the Lagrangian analytically  but let us
consider the following optimization
\begin{align}
&{\rm minimize\ \ \ }\qquad\qquad  \gamma \nonumber \\
&{\rm subject \ to}\quad  \gamma - \Sq(\rho, \bfx) = \mu(\bfx)
+\sum_{i} \lambda_i(\bfx) f_{\text{eq},i}(\bfx), \label{Eq:Higher
Order QCQP}
\end{align}
where each of the $\lambda_i$'s are polynomial functions of $\bfx$
and $\mu(\bfx)$ is a sum of squares polynomial (SOS) and therefore
nonnegative. That is $\mu(\bfx) = \sum_j [h_j(\bfx)]^2\geq 0$ for
some set of real polynomials $h_j(\bfx)$. The variables of the
optimizations are $\bfx$ and the coefficients that define the
polynomials $\mu(\bfx)$ and $\lambda_i$. Notice that we have $\gamma
\geq \Sq(\rho,\bfx) $ whenever the constraints are satisfied so that
once again we have a global upper bound on $\Sq(\rho,\bfx)$. This
optimization can be implemented numerically by restricting
$\mu(\bfx)$ and  $\lambda_i$ to be of some fixed degree. The
Lagrange dual optimization~\eqref{Eq:SDP-relaxation:deg2} arises
from choosing the degree of $\lambda_i$ to be zero. It is known that
for any fixed degree this optimization is an
SDP~\cite{P.A.Parrilo:2003,J.B.Lasserre:SJO:2001} and we have
implemented up to degree four using
SOSTOOLS~\cite{SOSTOOLS,S.Prajna:LNCIS:2005}. Schm\"{u}dgen's
theorem guarantees that by increasing the degree of the polynomials
in the relaxation we obtain bounds approaching the true maximum
$\Sq(\rho)$. This is a special case of the general procedure
described
in~\cite{P.A.Parrilo:2003,S.Prajna:LNCIS:2005,K.Schmudgen:MA:1991}
which is able to handle inequality constraints. For more details see
Appendix~\ref{sec:semidefinite relaxation}.

\section{Applications \& Limitations}\label{sec:examples}

In this section, we will look at some concrete examples of how the
two algorithms can be used to determine if some quantum states
violate a Bell inequality. Specifically, we begin by looking at how
the second algorithm can be used to determine, both numerically and
analytically, if some bipartite qudit state violates the Bell-CHSH
inequality. Then in Sec.~\ref{sec:Bell-3322}, we demonstrate how the
two algorithms can be used in tandem to determine if a class of
two-qubit states violate a Bell-3322
inequality~\cite{D.Collins:JPA:2004}. We will conclude this section
by pointing out some limitations of the UB algorithm that we have
observed.

\subsection{Bell-CHSH violation for Two-Qudit States}\label{sec:Bell-CHSH}

The Bell-CHSH inequality, as given by Eq.~\eqref{Eq:Bell-CHSH-inequality},
is one that amounts to choosing
\begin{equation}
b=\left(\begin{array}{cc}
1 & 1\\
1 & -1
\end{array}\right).\label{Eq:b-CHSH:dfn}
\end{equation}
For low-dimensional quantum systems, an upper bound on $\Sq(\rho)$
can be efficiently computed in MATLAB following the procedures
described in Sec.~\ref{sec:State-dependent Bound}. However, for
high-dimensional quantum systems, intensive computational resources
are required to compute this upper bound, which may render the
computation infeasible in practice. Fortunately, for a specific
class of two-qudit states, namely those whose coherence
vectors~\cite{fn:coherence vectors} vanish, it can be  shown that
(Appendix~\ref{Sec:UpperBound:Analytic}) their
$\Sq(\rho)=\max_{\Bell_\text{CHSH}}\eBellCHSH_\rho$ cannot exceed
\begin{equation}\label{Eq:CHSHViolation:UB}
\max_{z_1,z_2,z_3,z_4}
2\sqrt{2}s_1d\sqrt{\prod_{i=1}^2
\frac{2d^2-z_{2i-1}^2-z_{2i}^2}{2d^2}}+\sum_{k,l}b_{kl}\frac{z_kz_{l+2}}{d^2},
\end{equation}
where $s_1$ is the largest singular value of the matrix $R'$ defined
in Eq.~\eqref{Eq:R':Defn}, and $z_m$ is the trace of the dichotomic
observable $O_m$ given in Eq.~\eqref{Eq:z}.

To violate the Bell-CHSH inequality, we must have $\Sq(\rho)>2$,
hence for this class of quantum states, the Bell-CHSH inequality
cannot be violated if
\begin{equation}
\max_{z_1,z_2,z_3,z_4}\sqrt{2}s_1d
\sqrt{\prod_{i=1}^2\frac{2d^2-z_{2i-1}^2-z_{2i}^2}{2d^2}}
+\sum_{k,l}b_{kl}\frac{z_kz_{l+2}}{2d^2}\le
1.\label{Eq:CHSHViolation:Criterion}
\end{equation}
Since the Bell-CHSH inequality is tight
Eq.~\eqref{Eq:CHSHViolation:Criterion} guarantees the existence of a
LHV model for this experimental setup~\cite{A.Fine:PRL:1982}.
Essentially, the semianalytic upper bound
Eq.~\eqref{Eq:CHSHViolation:UB} was obtained by considering a
particular choice of Lagrange multipliers in the Lagrange dual
function~\eqref{Eq:Lagrange Dual}. Hence, it is generally not as
tight as the upper bound obtained numerically using the procedures
described in Sec.~\ref{sec:State-dependent Bound} (see
Appendix~\ref{Sec:UpperBound:Analytic} for details).

As an example, consider the $d$-dimensional isotropic state, i.e.,~a
mixture of the $d$-dimensional maximally entangled state
$\ket{\Psi^+_d}\equiv1/\sqrt{d}\sum_{j=1}^d\ket{j}\ket{j}$ and the
completely mixed state:
\begin{equation}
\rho_{I_d}=p\,\ketbra{\Psi^+_d}
+(1-p)\frac{\unit_{d^2}}{d^2}.\label{Eq:IsotropicState}
\end{equation}
As can be verified using the Positive Partial Transposition (PPT)
criterion~\cite{PPT,MPR.Horodecki:0109124}, this state is entangled
if and only if $p>p_{\rm Ent}\equiv1/(d+1)$. Using the procedures
outlined in Sec.~\ref{sec:State-dependent Bound}, we can numerically
compute, up till $d=5$, the threshold value of $p$ below which there
can be no violation of the Bell-CHSH inequality; these critical
values, denoted by $p_\text{UB-numerical}$ can be found in column 4
of Table~\ref{tbl:IsotropicStates:Bell-CHSH}. Similarly, we can
numerically compute the corresponding threshold values, denoted by
$p_\text{UB-semianalytic}$, using
Eq.~\eqref{Eq:CHSHViolation:Criterion}. It is worth noting that
these threshold values, as can be seen from column 3 and 4 of
Table~\ref{tbl:IsotropicStates:Bell-CHSH}, agree exceptionally well,
thereby suggesting that the computationally feasible criterion given
by Eq.~\eqref{Eq:CHSHViolation:Criterion} may be exact for the
isotropic states.

\begin{table}[!h]
\caption{\label{tbl:IsotropicStates:Bell-CHSH} The various threshold
values for isotropic states. The first column of the table is the
dimension of the local subsystem $d$. From the second column to the
fifth column, we have respectively the value of $p$ beyond which the
state is entangled $p_{\rm Ent}$, the value of $p$ below which
Eq.~\eqref{Eq:CHSHViolation:Criterion} is satisfied
$p_\text{UB-semianalytic}$, the value of $p$ below which the upper
bound obtained from lowest order relaxation is compatible with
Bell-CHSH inequality and the value of $p$ beyond which a Bell-CHSH
violation has been observed using the LB algorithm.}
\begin{ruledtabular}
\begin{tabular}{r|cccc}
$d$ & $p_{\rm Ent}$ & $p_\text{UB-semianalytic}$
& $p_\text{UB-numerical}$ & $p_\text{LB}$
\\ \hline
  2 &  0.33333 & 0.70711 & 0.70711 & 0.70711\\
  3 &  0.25000 & 0.70711 & 0.70711 & 0.76297\\
  4 &  0.20000 & 0.65465 & 0.65465 & 0.70711\\
  5 &  0.16667 & 0.63246 & 0.63246 & 0.74340\\
 10 &  0.09091 & 0.51450 & - & 0.70711 \\
 25 &  0.03846 & 0.36490 & - & 0.71516 \\
 50 &  0.01961 & 0.26963 & - & 0.70711 \\
\end{tabular}
\end{ruledtabular}
\end{table}

\subsection{$I_{3322}$-violation for Two-Qubit States}\label{sec:Bell-3322}

Next, we look at how the two algorithms can be used in tandem to
determine if some two-qubit states violates a Bell-3322 inequality,
in particular the $I_{3322}$ inequality~\cite{D.Collins:JPA:2004}.
This Bell inequality is interesting in that there are quantum states
that violate this new inequality but not the Bell-CH nor the Bell-CHSH
inequality. The analogue of Horodecki's criterion for this
inequality is thus very desired for the characterization of quantum
states that are inconsistent with local realistic description.

To the best of our knowledge, such an analytic criterion is yet to
be found. However, by combining the two algorithms presented above,
we can often offer a definitive, yet nontrivial, conclusion about
the compatibility of a quantum state description with that given by
LHV models. To begin with, we recall that the $I_{3322}$  inequality
reads:
\begin{gather}
\SLHV=p^{+-}_{AB}(1,1)+p^{+-}_{AB}(1,2)+p^{+-}_{AB}(1,3)+p^{+-}_{AB}(2,1)\nonumber\\
\:\quad\quad+p^{+-}_{AB}(2,2)-p^{+-}_{AB}(2,3)+p^{+-}_{AB}(3,1)-p^{+-}_{AB}(3,2)\nonumber\\
-p^+_A(1)-2p^-_B(1)-p^-_B(2)\le 0.\label{Eq:Bell-3322}
\end{gather}

Together with Eq.~\eqref{Eq:S:dfn} and
Eq.~\eqref{Eq:QM:probabilities}, one can then obtain the
corresponding Bell operator for probabilities:
\begin{align}\label{Eq:Bell-3322:Operator}
\mathcal{B}_{3322}=
& A^+_1\ten (B^-_1-B^+_2+B^-_3) -A^+_2\ten B^-_3\nonumber\\
-&A^-_2\ten (B^-_1+B^-_2) - A^+_3\ten B^-_2 - A^-_3 \ten
B^-_1,
\end{align}
where we have also used Eq.~\eqref{Eq:POVM:Normalization} to arrive
at this form. For convenience, we will adopt the notation that
$O^{\pm}_m\equiv A^{\pm}_m$ for $m=1,2,3$ and $O^{\pm}_m\equiv
B^{\pm}_{m-3}$ for $m=4,5,6$. In these notations, the global
optimization problem for this Bell inequality can be written as
\begin{subequations}\label{Eq:GlobalOptimization:3322}
\begin{gather}
\text{maximize \ \ } \tr(\rho~\Bell_{3322}) \label{Eq:eBell-3322}\\
\text{subject to \ \ }
\left(O^\pm_m\right)^2=O_m,\label{Eq:Constraints:Projector}
\end{gather}\end{subequations}
which is again a QCQP~\cite{fn:UB:Two-outcome}. The lowest order
relaxation to this problem can thus be obtained by following similar
procedures as that described in Sec.~\ref{sec:upper-bound}.

To obtain a {\em  state-dependent} upper bound on $\Sq(\rho)$ for
this inequality, we have to impose the analogue of
Eq.~\eqref{Eq:Range:z}, i.e.,
\begin{equation}
z^{\pm}_m=\tr(O^{\pm}_m)=0,1,\ldots,d,\label{Eq:Range:xpm0}
\end{equation}
for each of the POVM elements. For small $d$, numerical upper bounds
on $\Sq(\rho)$ can then be solved for
using SOSTOOLS. As an example, let's now look at how this upper
bound, together with the LB algorithm, has enabled us to determine
if a class of mixed two-qubit states violates the $I_{3322}$
inequality.

The mixed two-qubit state
\begin{gather}
\rho_{CG}=p\,\ketbra{\Psi_{2:1}}+(1-p)\,\Ketbra{0}{1},\quad 0\le
p\le 1,\label{Eq:State:CG}
\end{gather}
can be understood as a mixture of the pure product state
$\ket{0}\ket{1}$ and the nonmaximally entangled two-qubit state
$\ket{\Psi_{2:1}}=\frac{1}{\sqrt{5}}(2\ket{0}\ket{0}+\ket{1}\ket{1})$.
As can be easily verified using the PPT criterion~\cite{PPT}, this
state is entangled for $0<p\le 1$. In particular, the mixture with
$p=0.85$ was first presented in Ref.~\cite{D.Collins:JPA:2004} as an
example of a two-qubit state that violates the $I_{3322}$ inequality
but not the Bell-CH nor the Bell-CHSH inequality.

\begin{figure}[h!]
\includegraphics[width=8cm,height=2.29cm]{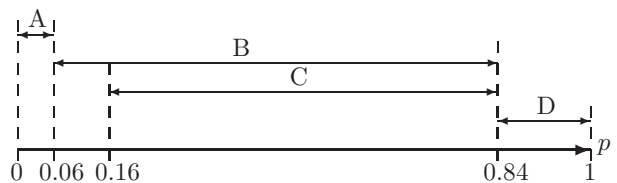}
\caption{\label{Fig:S:CG2:Bounds}Domains of $p$ where the
compatibility between local realism and quantum mechanical
description given by $\rho_{CG}$ were studied via the $I_{3322}$
inequality. From right to left are respectively the domain of $p$
whereby  $\rho_{CG}$ is: (D) found to violate the $I_{3322}$
inequality; (C) found to give a lowest order upper bound that is
compatible with the $I_{3322}$ inequality; (B) found to give a
higher order upper bound that is compatible with the $I_{3322}$
inequality; (A) not known if it violates the $I_{3322}$ inequality.}
\end{figure}

Given the above observation, a natural question that one can ask is,
at what values of $p$ does $\rho_{CG}$ violate the $I_{3322}$
inequality? Using the LB algorithm, we have found that for $p\gtrsim
0.83782$ (domain D in Fig.~\ref{Fig:S:CG2:Bounds}), $\rho_{CG}$
violates the $I_{3322}$ inequality. As we have pointed out in
Sec.~\ref{sec:lower-bound}, observables that lead to the observed
level of $I_{3322}$-violation can be readily read off from the output
of the SDP. 

On the other hand, through the
UB algorithm, we have also found that, with the lowest order
relaxation, the states do not violate this 3-setting inequality for
$0.16023 \lesssim p \lesssim 0.83625$ (domain C in
Fig.~\ref{Fig:S:CG2:Bounds}); with a higher order relaxation, this
range expands to $0.06291 \lesssim p \lesssim 0.83782$ (domain B in
Fig.~\ref{Fig:S:CG2:Bounds}); the next order relaxation is,
unfortunately, beyond what the software can handle. 

The algorithms alone therefore leave a tiny gap at $0 < p \lesssim
0.06291$ (domain A in Fig.~\ref{Fig:S:CG2:Bounds}) where we could not
conclude if $\rho_{CG}$ violates the $I_{3322}$
inequality. Nevertheless, if we recall that the set of quantum states
not violating a given Bell inequality is convex and that
$\rho_{CG}(p=0)$, being a pure product state, cannot violate any Bell
inequality, we can immediately conclude that $\rho_{CG}$ with $0\le p
\lesssim 0.83782$ cannot violate the $I_{3322}$ inequality. As such,
together with convexity arguments, the two algorithms allow us
to fully characterize the state $\rho_{CG}$ compatible with LHV theories,
when each observer is only allowed to perform three different dichotomic
measurements.

\subsection{Limitations of the UB algorithm}\label{sec:Limitation:UB}

As can be seen in the above examples, the UB algorithm does not
always provide a very good upper bound for $\Sq(\rho)$. In fact, it
has been observed that for pure product states, the algorithm with
lowest order relaxation always returns a state-independent bound
(the Tsirelson bound in the case of Bell-CHSH inequality). As such,
for mixed states that can be decomposed as a high-weight mixture of
pure product state and some other entangled state, the upper bound
given by UB is typically bad. To illustrate this, let us consider
the following 1-parameter PPT bound entangled
state~\cite{P.Horodecki:PLA:1997,MPR.Horodecki:PRL:1998}:\begin{subequations}\label{Eq:State:H3}
\begin{equation}
\rho_{H}=\frac{8p}{8p+1}\rho_\text{Ent}+\frac{1}{8p+1}\ketbra{\Psi_p},\quad
0< p <1
\end{equation}
where
\begin{gather}
\rho_\text{Ent}=\frac{1}{8}\sum_{i,j=0,i\neq
  j}^2\Ketbra{i}{j}-\frac{1}{8}\Ketbra{2}{0}\nonumber\\
+\frac{3}{8}\ket{\Psi^+_3}\bra{\Psi^+_3}\qquad\qquad\qquad\qquad\qquad\nonumber\\
\ket{\Psi_p}=\ket{2}\left(\sqrt{\frac{1+p}{2}}\ket{0}+\sqrt{\frac{1-p}{2}}\ket{2}\right).
\end{gather}\end{subequations}
It can be shown, for example, using the range
criterion~\cite{P.Horodecki:PLA:1997} that this
state is entangled when $0<p\le 1$. It is, however, well-known that
a bipartite PPT state cannot violate the Bell-CH nor the Bell-CHSH
inequality~\cite{R.Werner:PRA:2000}.

\begin{figure}[h!]
\includegraphics[width=8.5cm,height=6.6cm]{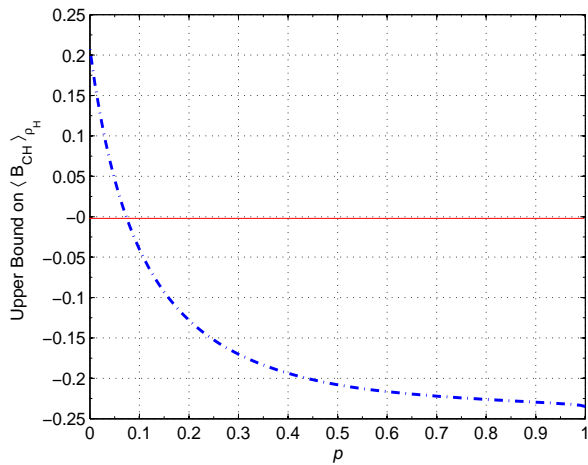}
\caption{\label{Fig:HBE} (Color online) Numerical upper bound on
maximal $\eBellCH_{\rho_{H}}$ obtained from the UB algorithm using
lowest order relaxation and Eq.~\eqref{Eq:Range:xpm0}. The dotted
horizontal line is the threshold above which no LHV description is
possible.}
\end{figure}

When tested with the UB algorithm using the lowest order relaxation,
it turned out that some of these upper bounds are actually above the
threshold of Bell-CH violation (see Fig.~\ref{Fig:HBE}). In fact,
the upper bound obtained for the pure product state,
$\rho_H(p=0)=\ketbra{\Psi_p}$ is actually the maximal achievable
Bell-CH violation given by a quantum
state~\cite{B.S.Tsirelson:LMP:1980}.

\section{Conclusion}

In this paper, we have presented two algorithms which can be used
to determine, respectively, a lower bound (LB) and an upper bound
(UB) on the maximal expectation value of a Bell operator for a given
quantum state, i.e., $\Sq(\rho)$. In particular, we have
demonstrated how one can make use of the upper bound to derive a
necessary condition for two-qudit states with vanishing coherence
vectors to violate the Bell-CHSH inequality.

For low dimensional quantum systems, we have also demonstrated how
one can make use of the two algorithms to determine, numerically, if
the quantum mechanical prediction is compatible with local realistic
description. On a separate note, these algorithms have also been
applied to the search of maximal-Bell-inequality-violation in the
context of collective measurements without
postselection~\cite{Y.C.Liang:2005b}.

As with many other numerical optimization algorithms, the algorithm
to determine a lower bound  (LB) on $\Sq(\rho)$, can only guarantee
the convergence to a local maximum. The UB algorithm, on the other
hand, provides an (often loose)  upper bound on $\Sq(\rho)$. In the
event that these bounds agree (up till reasonable numerical
precision), we know that optimization of the corresponding Bell
operator using LB has been achieved. This ideal scenario, however,
is not as common as we would like it to be. In particular, the UB
algorithm with lowest order relaxation has been observed to give
rather bad bounds for states with a high-weight mixture of pure
product states (although it appears that we can often rule out the
possibility of a violation in this situation by convexity arguments as
in Sec.~\ref{sec:Bell-3322}). A possibility to improve these bounds,
as suggested  by the work of Nie {\em et~al.}~\cite{J.Nie:MP:2006}, is
to incorporate the {\em Karush-Kuhn-Tucker} optimality condition as an
additional constraint to the problem. We have done some preliminary
studies on this but have not so far found any improvement in the
bounds obtained but this deserves further study.

As of now, we have only implemented the UB algorithm to determine
upper bounds on $\Sq(\rho)$ for dichotomic Bell inequalities. For
Bell inequalities with more outcomes, the local Hermitian
observables are generally also subjected to constraints in the form
of a LMI. Although the UB algorithm can still be implemented for
these Bell inequalities by first mapping the LMI to a series of
polynomial inequalities, this approach seems blatantly inefficient.
Future work to remedy this difficulty is certainly desirable.

Finally, despite the numerical and analytic evidence at hand, it is
still unclear why the lowest order relaxation to the global
optimization problem, as described in
Sec.~\ref{sec:State-independent Bound}, seems always gives rise to a
bound that is state-independent and how generally this is true. Some
further investigation on this may be useful, particularly to
determine whether the lowest order relaxation is always state
independent even for inequalities that are not correlation
inequalities. If so this could complement the methods of
Refs.~\cite{S.Wehner:PRA:2006,D.Avis:JPA:2006,M.Navascues:quant-ph:0607119}
for finding state-independent bounds on Bell inequalities.

\begin{acknowledgments}
This work was supported by the Australian Research Council. We thank
Ben Toner and Stephanie Wehner for fruitful discussions. Y.~-C.~Liang
would also like to thank Chris Foster and Shiang Yong Looi for their
suggestions on the actual implementation of the algorithms.
\end{acknowledgments}

\appendix

\section{Explicit forms of Semidefinite Programs}\label{sec:SDP}
Here, we provide an explicit form for the matrices $F_m$ and
constants $c_m$ that define the semidefinite
program~\eqref{Eq:sdpiter}. By setting
\begin{gather*}
Z=B^1_1\oplus\ldots \oplus B_1^{n_B}\oplus B_2^1\oplus\ldots\oplus
B^{n_B}_{m_B},\nonumber\\
F_0=-\left(\rho_B\,^1_1\oplus\ldots \oplus \rho_B\,_1^{n_B\,}\oplus
\rho_B\,_2^1\oplus\ldots\oplus \rho_B\,^{n_B}_{m_B}\right),
\end{gather*}
in Eq.~\eqref{Eq:SDP:Matrix}, we see that the inequality constraint
\eqref{Eq:SDP:Matrix:Ineq} of the SDP entails the positive
semidefiniteness of the POVM elements
$\left\{B^\lambda_l\right\}_{\lambda=1}^{n_B}$, and hence
Eq.~\eqref{Eq:SDPIter:ineq}. On the other hand, the equality
constraint \eqref{Eq:SDP:Matrix:Eq}, together with appropriate
choice of $F_m$ and $c_m$, ensures that the normalization condition
\eqref{Eq:SDPIter:eq} is satisfied.

In particular, each $F_m$ is formed from a direct sum of Hermitian
basis operators. A convenient choice of such basis operators is
given by the traceless Gell-Mann matrices, denoted by
$\{\sigma_i\}_{n=1}^{d^2-1}$, supplemented by
$\sigma_0=\unit_d/\sqrt{d}$ such that
\begin{equation}
\tr\left(\sigma_n\sigma_{n'}\right)=\delta_{nn'}\quad\text{and}\quad
\tr\left(\sigma_n\right)=\sqrt{d}\,\delta_{n0},\label{Eq:basis:Gell-Mann}
\end{equation}
where $d=d_B$ is the dimension of the state space that each
$B^\lambda_l$ acts on. A typical $F_m$ then consists of $n_B$
diagonal blocks of $\sigma_i$ at positions corresponding to the
$n_B$ POVM elements $\{B^\lambda_l\}_{\lambda=1}^{n_B}$ in $Z$ for a
fixed $l$. For instance, the set of $F_m$
\begin{equation*}
F_m=\left(
\begin{array}{cccccc}
\sigma_{m-1} & \zero & \zero & \zero  & \zero & \zero\\
\zero & \ddots & \zero & \zero & \zero  & \zero \\
\zero & \zero & \sigma_{m-1} & \mathbf{0}  & \zero & \zero\\
\zero & \zero & \mathbf{0} & \zero &\zero & \zero\\
\zero & \zero & \mathbf{0} & \zero &\ddots & \zero\\
\zero & \zero & \mathbf{0} & \zero &\zero & \zero
\end{array}\right),\quad 1\le m\le d^2,
\end{equation*}
together with $c_m=\sqrt{d}~\delta_{m1}$ entails the normalization
of $\{B^\lambda_1\}_{\lambda=1}^{n_B}$, i.e.,
$\sum_{\lambda=1}^{n_B} B^\lambda_1=\unit_{d_B}$; the remaining
$(m_B-1)d^2$ $F_m$ are defined similarly and can be obtained by
shifting the nonzero diagonal blocks diagonally downward by
appropriate multiples of $n_B$ blocks. The SDP thus consists of
solving Eq.~\eqref{Eq:SDP:Matrix} for a $m_Bn_Bd\times m_Bn_Bd $
Hermitian matrix $Z$ subjected to $d^2m_B$ affine constraints.

\section{Semidefinite Relaxation to the Global Optimization Problem}\label{sec:semidefinite relaxation}

The global optimization problem, either in the form of
Eq.~\eqref{Eq:GlobalOptimization:Correlation} for a two-outcome Bell
correlation inequality, or Eq.~\eqref{Eq:GlobalOptimization:3322} for a
two-outcome Bell inequality for probabilities, is a QCQP. As was
demonstrated in Sec.~\ref{sec:upper-bound}, an upper bound on
$\Sq(\rho)$ can then be obtained by considering the corresponding {\em
  Lagrange Dual}.

More generally, the global optimization problem can be mapped to a
real polynomial optimization
problem:\begin{subequations}\label{Eq:polynomial optimization}
\begin{align}
&\text{maximize \ } f_\text{obj}(\bfy),\label{Eq:Global.Obj}\\
&\text{subject to \ }
f_{{\rm eq},i}(\bfy)=0,\quad i=1,2,\ldots,N_{eq},\label{Eq:Global.Eq}\\
&\qquad\qquad\quad f_{{\rm ineq},j}(\bfy)\ge 0,\quad
j=1,2,\ldots,N_{ineq},\label{Eq:Global.Ineq}
\end{align}\end{subequations}
where $\bfy$ is a  vector of {\em real} variables formed by the
expansion coefficients of local observables $\{O_m\}$ in terms of
Hermitian basis operators.

Results from semialgebraic geometry dictate that an upper bound for
$f_\text{obj}(\bfy)$ can be computed  using Positivstellensatz-based
relaxations (see, for example, Ref.~\cite{P.A.Parrilo:2003} and
references therein). In particular, $\gamma$ will be an upper bound
on the constrained optimization problem \eqref{Eq:polynomial
optimization} if there exists a set of sum of squares (SOS)
$\mu_i(\bfy)$'s (i.e., nonnegative, real polynomials that can be
written as $\sum_j [h_j(\bfy)]^2$ with $h_j(\bfy)$ being some real
polynomials of $\bfy$), and  a set of real polynomials $\nu_j(\bfy)$
such
that~\cite{P.A.Parrilo:2003,S.Prajna:LNCIS:2005,K.Schmudgen:MA:1991}
\begin{align}
\gamma-f_\text{obj}(\bfy)=&\mu_0(\bfy)+\sum_j\nu_j(\bfy)f_{{\rm eq},j}(\bfy)+\sum_i\mu_i(\bfy)f_{{\rm
    ineq},i}(\bfy)\nonumber\\
\label{Eq:G(x)}&+\sum_{i_1,i_2}\mu_{i_1,i_2}(\bfy)f_{{\rm ineq},i_1}(\bfy)f_{{\rm
    ineq},i_2}(\bfy)+\ldots.
\end{align}
The relaxed optimization problem then consists of minimizing
$\gamma$ subjected to the above constraint. Clearly, at values of
$\bfy$ where the constraints are satisfied, $\gamma$ gives an upper
bound on $f_\text{obj}(\bfy)$. The auxiliary polynomials
$\nu_j(\bfy)$ and SOS $\mu_i(\bfy)$ then serve as the Lagrange
multipliers in the relaxed optimization problem.

For a fixed degree of the above expression, this relaxed
optimization problem can be cast as an SDP in the form of
Eq.~\eqref{Eq:SDP:Vector}~\cite{P.A.Parrilo:2003}. For the lowest
order relaxation, the auxiliary polynomials $\nu_j(\bfy)$ and SOS
$\mu_i(\bfx)$ are chosen such that degree of the expression in
Eq.~\eqref{Eq:G(x)} is no larger than the maximum degree of the set
of polynomials
\begin{equation*}
f_\text{obj}(\bfy),f_{{\rm eq},1}(\bfy),\ldots,f_{{\rm eq},N_{\rm
    eq}}(\bfy),f_{{\rm ineq},1}(\bfy), \ldots,f_{{\rm ineq},N_{\rm ineq}}(\bfy);
\end{equation*}
for a QCQP with no inequality constraints, this amounts to setting
all the $\mu_i(\bfy)$ to zero and all the $\nu_j(\bfy)$ to numbers.

For higher order relaxation, we increase the degree of the
expression in Eq.~$\eqref{Eq:G(x)}$ by increasing the degree of the
auxiliary polynomials. At the expense of involving more
computational resources, a tighter upper of $f_\text{obj}(\bfy)$ can
then be obtained by solving the corresponding SDPs.

\section{Lowest Order Relaxation with Observables of Fixed Trace}\label{sec:fixed-rank}

To obtain a tighter upper bound by using the lowest order relaxation
to \eqref{Eq:GlobalOptimization:Correlation}, we found it most
convenient to express \eqref{Eq:GlobalOptimization:Correlation} in
terms of the {\em real} optimization variables,
\begin{equation}\label{Eq:y:dfn}
y_{mn}\equiv \tr\left(O_m\sigma_n\right),\quad n=0,1,\ldots,d^2-1,
\end{equation}
which are just the expansion coefficients of each $O_m$ in terms of
a set of Hermitian basis operators satisfying
Eq.~\eqref{Eq:basis:Gell-Mann}. The constraints can then be taken
care of by setting each $y_{m0}=z_m/\sqrt{d}$. It is also expedient
to express the density matrix $\rho$ in terms of the same basis of
Hermitian operators
\begin{align*}
\rho=&\frac{\unit_{d^2}}{d^2}+\sum_{i=1}^{d^2-1}\left([\bfr_A]_i\sigma_i\ten\sigma_0
+[\bfr_B]_i\sigma_0\ten\sigma_i\right)\\
+&\sum_{i,j=1}^{d^2-1}[R']_{ij}\sigma_i\ten\sigma_j
\end{align*}
where
\begin{subequations}\label{Eq:rho:expansion}
\begin{gather}
[R']_{ij}=\tr\left(\rho~\sigma_i\ten\sigma_j\right),\label{Eq:R':Defn}\\
[\bfr_A]_i\equiv
\tr(\rho~\sigma_i\ten\sigma_0),\quad[\bfr_B]_j\equiv
\tr(\rho~\sigma_0\ten\sigma_j);\label{Eq:r:Defn}
\end{gather}\end{subequations}
$\bfr_A$, $\bfr_B$ are simply the coherence vectors that have been
studied in the literature~\cite{coherence vectors}.

We can thus incorporate the constraints \eqref{Eq:z} by expressing
the Lagrangian \eqref{Eq:Lagrangian} as a function of the reduced
set of variables
\begin{equation}
\bfy'^T\equiv[y_{11}~y_{12}~\ldots~y_{1\,d^2-1}~y_{21}~\ldots
y_{m_A+m_B\,d^2-1}],
\end{equation}
while all the $y_{m0}=z_m/\sqrt{d}$ are  treated as fixed parameters
of the problem. With this change in basis, and after some algebra,
the Lagrangian can be rewritten as
\begin{align}
\mathcal{L}(\bfy',\lambda_{mn})=&\sum_{m=1}^{m_A+m_B}\lambda_{m0}
\left(\sqrt{d}-\frac{z_m^2}{d\sqrt{d}}\right)+\sum_{k,l}b_{kl}\frac{z_kz_{l+m_A}}{d^2}\nonumber \\
&-\frac{1}{\sqrt{d}}({\bf
l}-\bfr)^T\bfy'-\bfy'^T\Omega'\bfy',\label{Eq:Lagrangian:FixedRank}
\end{align}
where $\lambda_{mn}$ are defined in Eq.~\eqref{Eq:Lambda:Expansion},
\begin{subequations}
\begin{gather}
\Omega'\equiv\half\left(
\begin{array}{cc}
\zero & -b\ten R' \\
-b^T\ten R'^T & \zero\\
\end{array}\right)+\bigoplus_{m=1}^{m_A+m_B} M_m,\nonumber\\
{\bf l}\equiv\vvec{L},\qquad  \bfr\equiv\left(\begin{array}{c}{\bf t}_A\ten\bfr_A\\
{\bf t}_B\ten\bfr_B\end{array}\right),\label{Eq:Omega,l,r:dfn}
\end{gather}
and for $i,j=1,2,\ldots,d^2-1$,
\begin{gather}
[L]_{jm}=2z_m\lambda_{mj},~[{\bf t}_A]_k=\sum_l
b_{kl}z_{l+m_A},~[{\bf t}_B]_l=\sum_k b_{kl}z_k,\nonumber\\
M_m=\sum_{n=0}^{d^2-1}\lambda_{mn}P_n,\quad
[P_n]_{ij}=\half\tr\left(\sigma_n\left[\sigma_i,\sigma_j\right]_+\right);
\end{gather}\end{subequations}
$[\sigma_i,\sigma_j]_+\equiv \sigma_i\sigma_j+\sigma_j\sigma_i$
is the anti-commutator of $\sigma_i$ and $\sigma_j$.

As before, we now maximize the Lagrangian
\eqref{Eq:Lagrangian:FixedRank} over $\bfy'$ to obtain the
corresponding Lagrange dual function. The latter, however, is
unbounded above unless
\begin{equation}
\left(
\begin{array}{cc}
-2t & \frac{1}{\sqrt{d}}\left({\bf l}^T-\bfr^T\right)\\
\frac{1}{\sqrt{d}}\left({\bf l}-\bfr\right) & 2\Omega'
\end{array}\right)\ge0, \label{Eq:Constraint:FixedRank}
\end{equation}
for some finite $t$. The convex optimization problem dual to
Eq.~\eqref{Eq:GlobalOptimization:Correlation} with fixed trace for
each observables is thus
\begin{align}
&\text{minimize}\sum_{m=1}^{m_A+m_B}\lambda_{m0}
\left(\sqrt{d}-\frac{z_m^2}{d\sqrt{d}}\right)+\sum_{k,l}b_{kl}\frac{z_kz_{l+m_A}}{d^2}
-t,\nonumber\\
&\text{subject to \ \ }\left(
\begin{array}{cc}
-2t & \frac{1}{\sqrt{d}}\left({\bf l}^T-\bfr^T\right)\\
\frac{1}{\sqrt{d}}\left({\bf l}-\bfr\right) & 2\Omega'
\end{array}\right)\ge0.\label{Eq:SDP:FixedRank}
\end{align}

\section{Nonlocality Criterion}\label{Sec:UpperBound:Analytic}

To derive the semianalytic criterion
Eq.~\eqref{Eq:CHSHViolation:Criterion}, we now note that any choices
of $\{\lambda_{mn}\}_{n=0}^{d^2-1}$ that satisfy constraint
\eqref{Eq:Constraint:FixedRank} will provide an upper bound on the
corresponding $\Sq(\rho)$. In particular, an upper bound can be
obtained by setting
\begin{equation}
\lambda_{mn}=\delta_{n0}\left[\lambda_A\left(\delta_{m1}+
\delta_{m2}\right)+\lambda_B\left(\delta_{m3}+\delta_{m4}\right)\right]
\end{equation}
and solving for $\lambda_A$, $\lambda_B$ that satisfy the constraint
\eqref{Eq:Constraint:FixedRank}. With this choice of the Lagrange
multipliers, and for quantum states with vanishing coherence vectors,
the constraint \eqref{Eq:Constraint:FixedRank} becomes
\begin{align}
\left(\begin{array}{ccc}
-2t & \zero & \zero \\
\zero & \frac{2\lambda_A}{\sqrt{d}}\unit_2\ten\unit_{d^2-1} & -b\ten R' \\
\zero & -b\ten R^{'T} & \frac{2\lambda_B}{\sqrt{d}}\unit_2\ten
\unit_{d^2-1}
\end{array}\right)\ge 0.
\end{align}
where $b$ and $R'$ are defined, respectively, in
Eq.~\eqref{Eq:b-CHSH:dfn} and Eq.~\eqref{Eq:R':Defn}. This, in turn is
equivalent to\begin{subequations}
\begin{gather}
-t\ge0,\label{Eq:Constraint:FixedRank:t}\\
\left(\begin{array}{cc}
\frac{2\lambda_A}{\sqrt{d}}\unit_2\ten\unit_{d^2-1} & -b\ten R'\\
-b\ten R^{'T} & \frac{2\lambda_B}{\sqrt{d}}\unit_2\ten \unit_{d^2-1}
\end{array}\right)\ge 0.\label{Eq:Constraint:FixedRank:Reduced}
\end{gather}\end{subequations}
Using Schur's complement~\cite{LA}, the
constraint~\eqref{Eq:Constraint:FixedRank:Reduced} can be explicitly
solved to give
\begin{equation*}
\lambda_A\lambda_B\ge \half s_1^2 d,
\end{equation*}
where $s_1$ is the largest singular value of the matrix $R'$.
Substituting this and Eq.~\eqref{Eq:Constraint:FixedRank:t} into
Eq.~\eqref{Eq:SDP:FixedRank}, and after some algebra, we see that
$\eBellCHSH_\rho$ for a quantum state $\rho$ with vanishing
coherence vectors cannot be greater  than
\begin{equation*}
\max_{z_1,z_2,z_3,z_4}
2\sqrt{2}s_1d\sqrt{\prod_{i=1}^2\frac{2d^2-z_{2i-1}^2-z_{2i}^2}{2d^2}}+\sum_{k,l}b_{kl}\frac{z_kz_{l+2}}{d^2}.
\end{equation*}
For $\rho$ to violate the Bell-CHSH inequality, we must have this
upper bound greater than the classical threshold value of 2 [c.f.
Eq.~\eqref{Eq:Bell-CHSH-inequality}]. Hence a sufficient condition
for $\rho$ to satisfy the Bell-CHSH inequality is given by
Eq.~\eqref{Eq:CHSHViolation:Criterion}.


\begin{thebibliography}{99}
\bibitem{fn:tight inequality} For a given experimental setup, the set
  of local realistic correlations defines a convex polytope; any facet
  of this polytope is a tight Bell inequality for the given
  experimental setup.
\bibitem{fn:probability-correlation} For a Bell correlation
inequality, we can apply LB by first rewriting the corresponding
Bell operator \eqref{Eq:BellOperator:Correlation1} in terms of POVM
elements that form the measurement outcomes of each $O_m$.
\bibitem{fn:Dichotomic:Defn} A dichotomic observable is a Hermitian
observable with only two distinct eigenvalues.
\bibitem{fn:CH-CHSH} A quantum state violates the Bell-CH inequality
  if and only if it violates the Bell-CHSH inequality.
\bibitem{fn:Bell-correlation} Strictly speaking, a general Bell
correlation operator may also contain marginal terms like
$O_m\ten\unit_{d_B}$ and $\unit_{d_A}\ten O_m$. For brevity, we will
not consider such Bell operators in the following discussion. It
should, nevertheless, be clear to the readers that the following
arguments are readily generalized to include Bell operators of this
more general kind.
\bibitem{fn:pm1} The choice of eigenvalues, nonetheless, has to be
consistent with the values assigned to the two outcomes of the
respective dichotomic Bell experiments.
\bibitem{fn:subsystems.diff.dim} In general, the composite system
may consist of subsystems of different dimensions. All the following
arguments can be readily generalized to this more general scenario.
\bibitem{fn:coherence vectors} These are generalization of the Bloch
vectors representation for higher dimensional quantum systems. See
also~\cite{coherence vectors}.
\bibitem{fn:UB:Two-outcome} As was mentioned in
  Sec.~\ref{sec:GlobalOptimizationProblem}, the same remark applies to
  any other  bipartite Bell-inequality for probabilities with only two
  outcomes   per measurement setting.
\end{thebibliography}
\end{document}